\documentclass[manuscript]{aastex}

\shorttitle{Jet Model for SN 2007gr}
\shortauthors{Xu et al.}

\begin{document}


\title{An Off-Axis Relativistic Jet Model for the Type Ic supernova SN 2007gr}


\author{M. Xu\altaffilmark{1,2,3}, S. Nagataki\altaffilmark{2}, and Y. F. Huang\altaffilmark{1,3}}
\altaffiltext{1}{Department of Astronomy, Nanjing University, Nanjing 210093, China; hyf@nju.edu.cn}
\altaffiltext{2}{Yukawa Institute for Theoretical Physics, Oiwake-cho, Kitashirakawa,
Sakyo-ku, Kyoto 606-8502, Japan}
\altaffiltext{3}{Key Laboratory of Modern Astronomy and Astrophysics (Nanjing University),
Ministry of Education, China}


\begin{abstract}
We propose an off-axis relativistic jet model for the Type Ic supernova SN 2007gr. Most of the energy
($\sim2\times10^{51}$ erg) in the explosion is contained in non-relativistic ejecta which
produces the supernova. The optical emission is coming from the decay process of $\rm ^{56}Ni$
synthesized in the bulk SN ejecta.
Only very little energy ($\sim10^{48}$ erg) is contained
in the relativistic jet with initial velocity about $0.94$ times the speed of
light.
The radio and X-ray emission
comes from this relativistic jet. With some typical parameters of a Wolf-Rayet star
(progenitor of Type Ic SN), i.e., the mass loss rate
$\dot{M}=1.0\times10^{-5}~M_{\odot}~\rm yr^{-1}$ and the
wind velocity $v_{\rm w}=1.5\times10^{3}~\rm km~s^{-1}$ together with an observing
angle of $\theta_{\rm obs} = 63.3^{\circ}$, we can obtain the multiband light curves that
fit the observations well. All the observed data are consistent with
our model. Thus we conclude that SN 2007gr contains a weak
relativistic jet and we are observing the jet from off-axis.
\end{abstract}

\keywords{gamma-ray bursts: general - supernovae: individual (SN 2007gr)}

\section{Introduction}

Gamma-ray bursts (GRBs) are intense flashes of gamma-ray radiation in the Universe
(for recent reviews, see: Zhang 2007; Gehrels et al. 2009). It is widely believed that
outflows of GRBs are accelerated to ultra-relativistic speeds (M\'{e}sz\'{a}ros 2002) and
usually collimated with small jet anlges (Frail et al. 2001).
 On the other side, core-collapse supernovae (SNe) are the
explosive deaths of massive stars that occur when their iron cores collapse to form neutron
stars or black holes (Wilson 1971; Barkat et al. 1974; Wheeler \& Levreault 1985;
Woosley \& Janka 2005; Woosley \& Bloom 2006;
Nagataki et al. 2007; Nagataki 2009; Nagataki 2010). According to their
spectra, core-collapse SNe are classified as Type Ib, Ic or Type II SNe
(Wheeler 1993; Wheeler et al. 1993; Filippenko 1997).
SN explosions will release massive ejecta and usually they are isotropic and non-relativistic.

Recent observations have revealed that many nearby GRBs are associated with core-collapse SNe.
Examples of such association includes GRB 980425/SN 1998bw (Galama et al. 1998),
GRB 030329/SN 2003dh (Berger et al. 2003),
GRB 031203/SN 2003lw (Cobb et al. 2004), GRB 060618/SN 2006aj (Campana et al. 2006),
GRB 091127/SN 2009nz (Cobb et al. 2010), and GRB 100316D/SN 2010bh (Fan et al. 2010) etc.
These GRBs are usually soft in $\gamma$-ray spectra and are ubiquitously longer than $2$ seconds.
Thus they belong to the so called long/soft GRBs.

The observed GRB-connected SNe are all Type Ic SNe.
The most favored progenitors for Type Ic supernovae are Wolf-Rayet stars
(Maeder \& Lequeux 1982; Begelman \& Sarazin 1986; Woosley \& Bloom 2006).
However, the GRB-connected Type Ic SNe should be
different from ordinary Type Ic SNe (Soderberg et al. 2006), because they need to
launch relativistic jets to produce the bursts of $\gamma$-rays.
The kinetic energy of these SNe  appears to be greater than that of ordinary
SNe. In some SNe associated with GRBs, most of the explosion energy is in non-relativistic ejecta which
produces the supernova, while only little energy is in the relativistic jets
which are responsible for making GRBs and their afterglows (Woosley \& Bloom 2006).

While GRB-connected SN explosions can produce relativistic
jets with Lorentz factor as large as $\sim 100$,
we believe that there were still some Type Ic SNe that could only produce
midly relativistic jets with initial Lorentz factor
of a few (Huang et al. 2002; Granot \& Loeb 2003). It has been argued that this kind of low
Lorentz factor jets will produce UV or soft X-ray transients
but not GRBs (Huang et al. 2002; Xu et al. 2010).
The interesting Type Ic supernova SN 2009bb, which is identified
as a relativistic ejecta without a detected GRB (Soderberg et al. 2010a),
may be such an event.

Another event, the Type Ic supernova SN 2007gr, is much more controversial.
Soderberg et al. (2010b) proposed that SN 2007gr is an ordinary
Type Ic supernova with an almost constant expansion speed ($v\propto t^{-0.1}$).
On the contrary, Paragi et al.'s (2010) 5 GHz Radio observations have revealed a
relativistic jet in SN 2007gr. While the opening angle of the jet
is similar to that of a typical GRB jet, its Lorentz factor seems to be
far smaller than a normal GRB outflow.
In view of the non-relativistic expansion of the photosphere of SN 2007gr
(Valenti et al. 2008; Hunter et al. 2009), here, we propose an
off-axis relativistic jet model for SN 2007gr with typical parameters
of circumstellar medium (CSM) of a Wolf-Rayet star. Most of the energy
in the explosion is contained in non-relativistic ejecta which
produces the supernova, while only a small fraction of expansion energy is contained
in the relativistic jet.
We first describe our dynamical model in Section 2. In section 3 we describe the
parameters used in our modeling. The model results are shown in Section 4. Our
conclusions and discussion are presented in Section 5.

\section{Model}

On 2007 August 15.51 UT, SN 2007gr was discovered by the Katzmann Automatic
Imaging Telescope in NGC 1058 (Madison \& Li 2007), a bright spiral
galaxy belonging to a group of galaxies. The distance of one member of this group, NGC 935, has been
derived as about 9.3 Mpc (Silbermann et al. 1996). In our study, we adopt this value for
the distance of SN 2007gr. The explosion date was suggested as 2007 August $13\pm2$
(Soderberg et al. 2010b). Radio observations with the Very Large Array
(Soderberg et al. 2010b) and European VLBI network (EVN) (Paragi et al. 2010)
revealed a radio source in the place of the supernova.
SN 2007gr was classified as Type Ic supernova according to its spectra (Crockett et al.
2008; Valenti et al. 2008).

\subsection{Relativistic jet}

Paragi et al. (2010) reported the evidence for the existence of a relativistic
jet in 2007gr (with expansion speed $v>0.6 \rm c$) based on their radio observations,
while Valenti et al. (2008) and Hunter et al. (2009)
measured a non-relativistic velocity
(from $\sim11000~\rm km/s$ at 1 week after explosion to
$\sim4800~\rm km/s$ at 50 days after explosion)
for the photospheric expansion of SN 2007gr. These observations suggest that the
average speed of the optical
ejecta (i.e., optical expansion velocity) is non-relativistic while the radio ejecta
is much faster and relativistic. This scenario is similar to that of Soderberg et al. (2005,2010b),
but note that the velocity of the radio ejecta here is relativistic.
So, we suggest that SN 2007gr may phenomenologically contain two components:
a non-relativistic component and a relativistic component.
The non-relativistic component should contain most of the explosion energy and account for
the photospheric expansion and the optical emission. The relativistic component should be
a jet that contains only a small fraction of the explosion energy and account for the radio emission.
It is more likely central engine driven. Its behavior should be
similar to a GRB jet, but note that its initial Lorentz factor is significantly smaller.
In this paper, we will simulate the evolution of the relativistic
jet numerically and compare the results with observations.

In our framework, the optical emission of supernova should mainly come from the decay process of
$\rm ^{56}Ni$ synthesized in the SN explosion.
On the contrary, radio and X-ray emission of supernova is explained as synchrotron radiation from
relativistic electrons which are accelerated by the shock produced in the
collision between the jet and circumstellar medium (Chevalier 1998;
Soderberg et al. 2005). The shock process is very similar to the external shock process
of GRBs that gives birth to GRB afterglows.

\subsection{Jet dynamics}

The dynamical evolution of
a relativistic jet that collides with surrounding medium can be conveniently
described by the equations proposed by Huang et al. (1999, 2000).
Their method can be widely used in both
ultra-relativistic and non-relativistic phases.
In this study, we will adopt Huang et al.'s equations to simulate
the evolution of the external shock.
The evolution of the bulk Lorentz of the jet ($\gamma$), the swept-up mass of medium ($m$),
the radius of the shock ($R$), and the half-opening angle of the jet ($\theta$) are described
by the following equations,

\begin{equation}
\frac{d\gamma}{dm}=-\frac{\gamma^{2}-1}{M_{\rm ej}+\varepsilon m+2(1-\varepsilon)\gamma m},
\end{equation}

\begin{equation}
\frac{dm}{dR}=2\pi R^{2}(1- \cos \theta)n m_{\rm p},
\end{equation}

\begin{equation}
\frac{dR}{dt}=\beta c \gamma(\gamma+\sqrt{\gamma^2-1}),
\end{equation}

\begin{equation}
\frac{d\theta}{dt}=\frac{c_{\rm s}(\gamma+\sqrt{\gamma^2-1})}{R},
\end{equation}
where $M_{\rm ej}$ is the initial mass of the ejecta, $m_{p}$ is the mass of the proton,
and $\beta=\sqrt{\gamma^2-1}/\gamma$. The radiative efficiency ($\varepsilon$) is assumed
as zero because the ejecta becomes adiabatic a few hours after the burst.

In the above equations, the velocity of
the lateral expansion has been assumed to be the sound speed $c_{\rm s}$,
which can be calculated from  (Dai et al. 1999)
\begin{equation}
c_{\rm s}^2 = \frac{\hat{\gamma}(\hat{\gamma}-1)(\gamma-1)c^{2}}{1+\hat{\gamma}(\gamma-1)},
\end{equation}
where $\hat{\gamma}\approx(4\gamma+1)/(3\gamma)$ is the adiabatic index.
The number density of the circumstellar medium ($n$) is inversely proportional to
the square of the shock radius, i.e.
\begin{equation}
n=\frac{\dot{M}}{4\pi m_{\rm p}v_{\rm w}R^{2}},
\end{equation}
where $\dot{M}$ is mass loss rate of the circumstellar wind, and $v_{\rm w}$ is the wind speed.

\subsection{Synchrotron radiation process}

As usual, in the comoving frame, the shock-accelerated electrons are assumed to follow
a power-law distribution according to their Lorentz factors ($\gamma_{e}$)
\begin{equation}
\frac{dN_{\rm e}'}{d\gamma_{\rm e}}=(\gamma_{\rm e}-1)^{-p},
\end{equation}
where $p$ is the power-law index. Note that in the bracket, 1 is subtracted from $\gamma_{e}$ to account
for the non-relativistic phase (Huang \& Cheng 2003).
For a single electron with a Lorentz factor of $\gamma_{e}$,
the synchrotron radiation power at frequency $\nu'$ is given by
\begin{equation}
P(\nu',\gamma_{\rm e})=\frac{\sqrt{3}e^3B'}{m_{\rm e}c^2}F(\frac{\nu'}{\nu_{\rm c}'}),
\end{equation}
where $e$ is the electron charge, $B'$ is the  magnetic intensity, $m_{\rm e}$ is the mass of electron,
$\nu_c'=3\gamma_{\rm e}^{2}eB'/(4\pi m_{\rm e} c)$ and the function $F(x)$ is defined as
\begin{equation}
F(x)=x\int_{x}^{+\infty}K_{5/3}(k)dk,
\end{equation}
with $K_{5/3}(k)$ being the Bessel function.

The magnetic energy density is assumed to be a fraction $\epsilon_{B}^2$ of the total thermal energy density,
i.e. (Dai et al. 1999),
\begin{equation}
\frac{B'^2}{8\pi}=\epsilon_{\rm B}^2 \frac{\hat{\gamma}\gamma+1}{\hat{\gamma}-1}(\gamma-1)nm_{\rm p}c^2.
\end{equation}
Therefore, the total synchrotron radiation power from all the shock
accelerated electrons is
\begin{equation}
P(\nu')=\int_{\gamma_{\rm e,min}}^{\gamma_{\rm e,max}} \frac{dN_{\rm e}'}{d\gamma_{\rm e}}
P(\nu',\gamma_{\rm e})d\gamma_{\rm e},
\end{equation}
where $\gamma_{\rm e,max}=10^8(B'/1~\rm G)^{1/2}$ is the maximum Lorentz factor of elections,
$\gamma_{\rm e,min}=\epsilon_{\rm e}(\gamma-1)m_{\rm p}(p-2)/[m_{\rm e}(p-1)]+1$ is
the minimum Lorentz factor of elections,
and $\epsilon_e$ is electron energy faction.
Then we can obtain the observed flux density at frequency $\nu$,
\begin{equation}
F_{\nu}=\frac{1}{\gamma^{3}(1-\beta \cos\Theta)^{3}} \frac{1}{4\pi D_{L}^{2}}
   P'[\gamma(1-\beta \cos\Theta)\nu],
\end{equation}
where $D_{L}$ is the luminosity distance, and $\Theta$ is the angle between the line of sight
and the velocity of emitting material.

The synchrotron self absorption effect (Rybicki \& Lightman 1979) need to be
 considered in calculating the radio flux
(Chevalier 1998; Kong et al. 2009).
Self-absorption reduces the synchrotron radiation flux by a factor of $(1-e^{\tau_{\nu}})/\tau_{\nu}$,
where $\tau_{\nu}$ is the optical depth. The self-absorption coefficient is given by
\begin{equation}
k(\nu')=\frac{p+2}{8\pi m_{\rm e} \nu'^2}\int_{\gamma_{\rm e,min}}^{\gamma_{\rm e,max}} \frac{dN_{\rm e}'}{d\gamma_{\rm e}}
\frac{1}{\gamma_{\rm e}}P(\nu',\gamma_{\rm e})d\gamma_{\rm e}.
\end{equation}

When calculating the observed flux, we integrate the emission over the
whole equal arrival time surface (Waxman 1997; Sari 1997; Panaitescu \& M\'{e}sz\'{a}ros 1998)
determined by
\begin{equation}
t=\int\frac{1-\beta \Theta}{\beta c}dR\equiv {\rm const.}
\end{equation}

\section{Parameters}

Before fitting the observations of SN 2007gr numerically, we first give a rough
estimation of several fundamental parameters of the jet based on theoretical
analysis of some observed facts.

(i) The initial bulk Lorentz factor $\gamma_0$.
Paragi et al. (2010) have derived a conservative lower limit of 1.7 mas for the angular
diameter from their observations made with the EVN and the
Green Bank Telescope during 2007 November 5 --- 6 ($\sim85$ days after the supernova
explosion). This will correspond to an average expansion speed faster than 0.6 times the speed of light
($\beta >0.6$). In view of the deceleration of the ejecta, we assume that the initial
bulk Lorentz factor is 3, i.e., $\beta_0 \approx 0.94$.

(ii) The initial half-opening angle $\theta_0$.
The typical half-opening angle of GRB jets is about $0.1$ rad (Zhang 2007; Gao \& Dai 2010),
while supernova outflows are much more isotropic.
Since the Lorentz factor of the jet involved in SN 2007gr is much smaller than a typical GRB jet,
its opening angle should be correspondingly much larger.
Here we assume that the initial half-opening angle of the jet in our model
is $\theta_0 =0.6$ rad ($\sim34.4^{\circ}$).

(iii) The observing angle $\theta_{\rm obs}$, i.e. the angle between the line of sight and
the symmetry axis of the jet.
The ten hours of EVN observations of SN 2007gr during 2007 November 5 --- 6 restored an
elliptical beam with the size of $15.26\times6.85$ mas (Paragi et al. 2010).
In view of the none detection of prompt emission in the early stage of SN 2007gr, we suggest
that the jet is not pointed toward us, but off-axis. In this paper, the inclination angle between
the jet axis and our line of sight is taken as $\theta_{\rm obs}=1.1~\rm rad$ ($\sim63.3^{\circ}$).
Such an observing angle, together with the smaller initial
half-opening angle of $\theta_0=0.6~\rm rad$ ($\sim34.4^{\circ}$), means that our line of sight
is completely outside the jet boundry, i.e. we are observing the mildly relativistic jet
off-axisly.

(iv) The initial kinetic energy of the shock $E$.
The total kinetic energy and ejected mass of the SN ejecta is $\sim1.5-3\times10^{51}$ erg
and $\sim1.5-3.5~ M_{\odot}$ respectively (Valenti et al. 2008; Hunter et al. 2009).
The total internal energy for the observed radio emission is $\sim0.7-4.5\times10^{46}$ erg
(Soderberg et al. 2010b; Paragi et al. 2010). Because most of the energy in the explosion is
contained in non-relativistic ejecta and only very little energy is contained
in the relativistic jet, we assume that the initial kinetic energy of the relativistic
shock wave is $E=1.1\times10^{48}$ erg.

(v) The number density of the circumstellar medium ($n$).
SN 2007gr is classified as Type Ic SN according to its spectra
(Crockett et al. 2008; Valenti et al. 2008).
The most favored progenitors for Type Ic supernovae are
Wolf-Rayet stars (Maeder \& Lequeux 1982; Begelman \& Sarazin 1986; Woosley \& Bloom 2006).
For Wolf-Rayet stars, the typical mass loss rate is
$\dot{M}\sim0.6-17.1\times10^{-5}~M_{\odot}~ \rm yr^{-1}$,
and the wind speed is typically
$v_{\rm w}\sim0.7-5.5\times10^{3}~ \rm km~s^{-1}$ (Eenens \& Williams 1994; Cappa et al. 2004).
In our calculations, we assume the following parameters for the progenitor of SN 2007gr, i.e.,
$\dot{M}=1.0\times10^{-5}~M_{\odot}~\rm yr^{-1}$ and
$v_{\rm w}\sim1.5\times10^{3}~\rm km~s^{-1}$.
Then the number density of CSM can be calculated from Eq. (6).
The mass loss rate in our modeling is more than 10 times higher than the ones
in previous studies (Paragi et al. (2010); Soderberg et al. (2010b)).

(vi) The electron energy faction $\epsilon_{\rm e}$.
In our model, we assume an evolving electron energy faction as
$\epsilon_{\rm e}=\epsilon_{\rm e,0}\times(R/R_{0})^{\alpha}$.
The parameter $\epsilon_{\rm e}$ is evolving with time. It is
slightly different from a normal GRB model.
In our calculations, we assume
$\epsilon_{\rm e,0}=0.1$, $R_{0}=3.0\times10^{16}$ cm, $\alpha=5/4$, i.e.,
\begin{equation}
\epsilon_{\rm e}=0.1\times(\frac{R}{3.0\times10^{16}~ \rm cm})^{5/4}.
\end{equation}
Note that our $\epsilon_{\rm e}$ is smaller than unity during all the observing time.

For other parameters such as the magnetic energy fraction ($\epsilon_{\rm B}$)
and the power-law index of the energy distribution function of electrons ($p$),
we take $\epsilon_{\rm B} = 0.1$ and $p = 3.3$, respectively,
which are similar to those adopted by Soderberg et al. (2010b).

\section{Results}

Using our off-axis relativistic jet model and the parameters above, we have
numerically calculated the multi-band emission of the jet in
SN 2007gr. Here we compare the theoretical light curves with observations.

Fig. 1 shows the light curves of radio emission. From this figure, we see that the theoretical
curves at 1.4 GHz (Fig. 1a), 4.9 GHz (Fig. 1b), and 8.5 GHz (Fig. 1c) can fit the observational data well
(note the last data point in Fig. 1c is an upper limit).
The curve at 22.5 GHz (Fig. 1d) is also consistent with the upper limit of the observation.

\begin{figure}
\epsscale{1.0}
\plotone{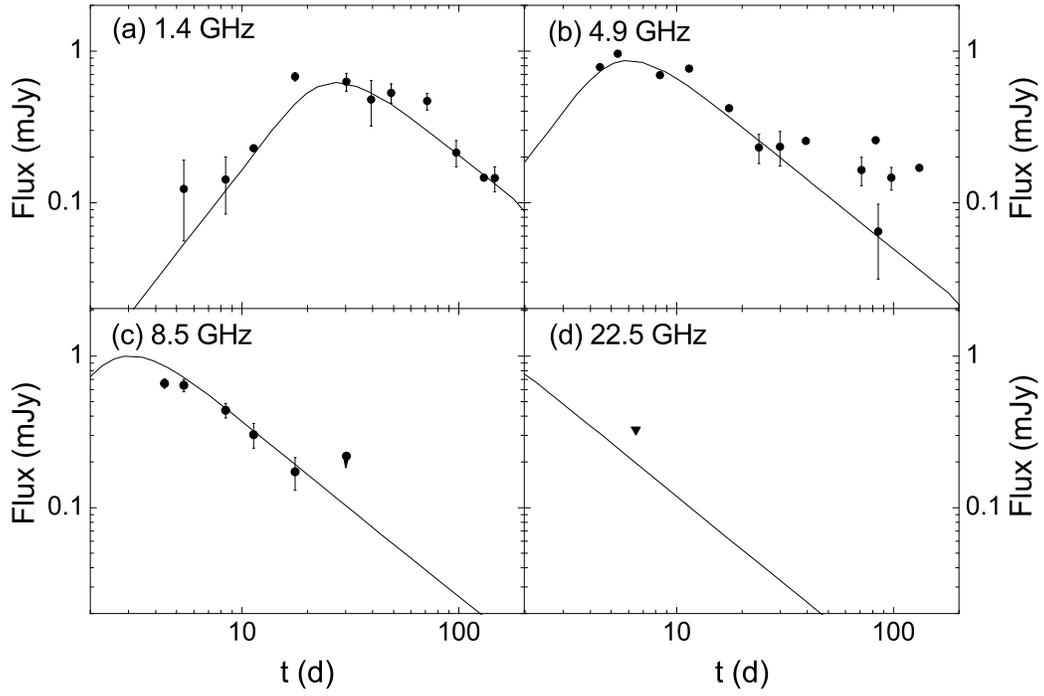}
\caption{Observed radio light curves of SN~2007gr and our best fit by using
the off-axis relativistic jet model. Panels (a), (b), (c), and (d) are light curves at 1.4, 4.9,
8.5 and 22.5 GHz respectively. Observed data points are taken from Soderberg et al. (2010b).
\label{fig1}
}
\end{figure}

Fig. 2 illustrates the theoretical optical and X-ray light curves based on our model.
Our X-ray light curve is consistent with the upper limit of the $Chandra$
observation (Soderberg 2007). The predicted optical emission
is significantly lower than the observed flux. This is a reasonable result.
We believe that the observed optical emission should mainly come from the
decay process of $\rm ^{56}Ni$ synthesized in the bulk SN ejecta (Arnett 1982; Sutherland \& Wheeler 1984),
as that typically happens in usual SN explosions.

\begin{figure}
\epsscale{1.0}
\plotone{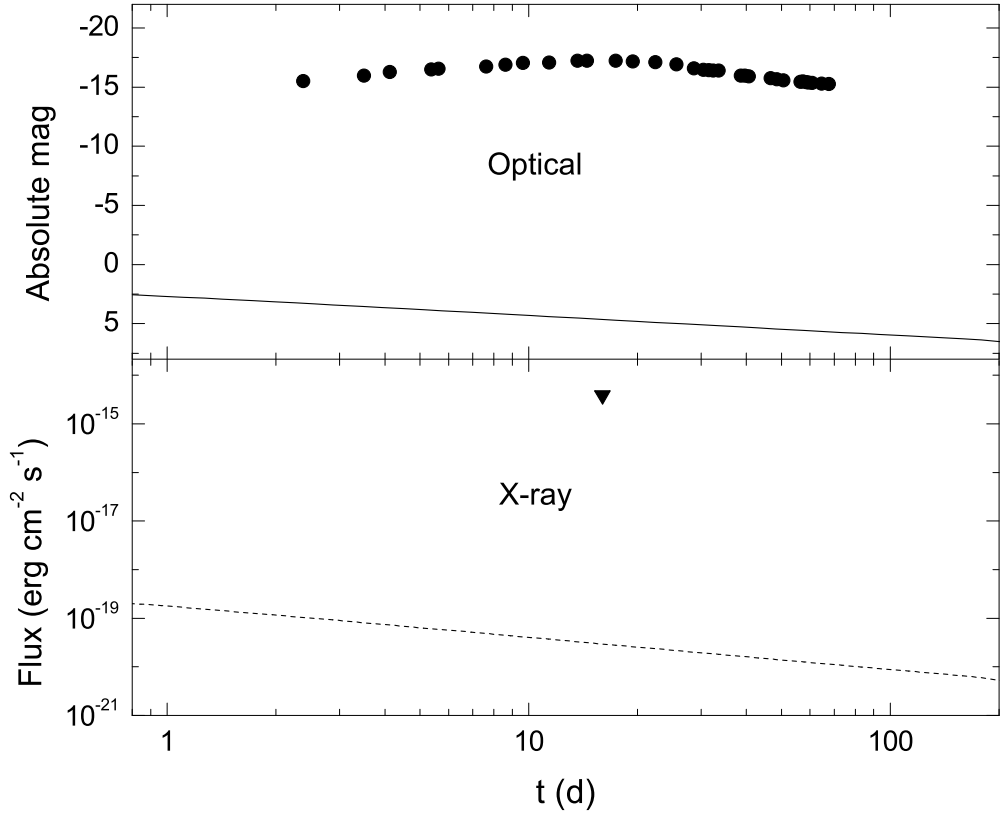}
\caption{Light curves of SN 2007gr in optical and X-ray bands. The solid and dashed curves
correspond to optical and X-ray emission from our off-axis relatisvistic jet, respectively.
The circular points are the observed R-band data taken from Valenti et al. (2008).
The triangle point is the upper limit observed by $Chandra~X-ray~Observatory$ (Soderberg 2007).
\label{fig2}}
\end{figure}

Figs. 1 and 2 are our best fits to the observations by adopting optimal values for
the parameters involved. To get these optimal parameters, we actually have tried many times.
In Figs. 3 and 4, we go further to give some examples of the radio light curves to illustrate
the effects of various parameters.
In these figures, the observed 1.4GHz
data points are taken from Soderberg et al. (2010b).
The solid curve corresponds to our best fit by using the
parameters described in Section 3. The dashed and dotted curves
are drawn with only one parameter altered.
From these figures, we see that the theoretical radio fluxes
depend sensitively on the parameters of $\theta$, $\dot{M}$, and $v_{\rm w}$.
The flux of a more isotropic outflow is higher
than that of a narrower outflow.
A fast wind with a low mass loss rate tends to increase the emission.
The parameters $\gamma,~E, ~\epsilon_{\rm B}$, and $\epsilon_{\rm e}$
affect not only the intensity, but also the shape and peak time of the light curve.
It means that the radiation spectra depends on these parameters sensitively.

As is shown in Fig. 4b, the light curve is flattened by a positive
$\alpha$. If $\epsilon_{\rm e}$ is constant
as normally assumed in GRB modeling, i.e., $\alpha=0$, then the flux decreases much
faster than the positive $\alpha$ scenario.
A positive $\alpha$ is necessary in our fitting, which indicates that
$\epsilon_{\rm e}$ is increasing with time, i.e., more and more kinetic energy is
gradually transformed to internal energy.

\begin{figure}
\epsscale{1.0}
\plotone{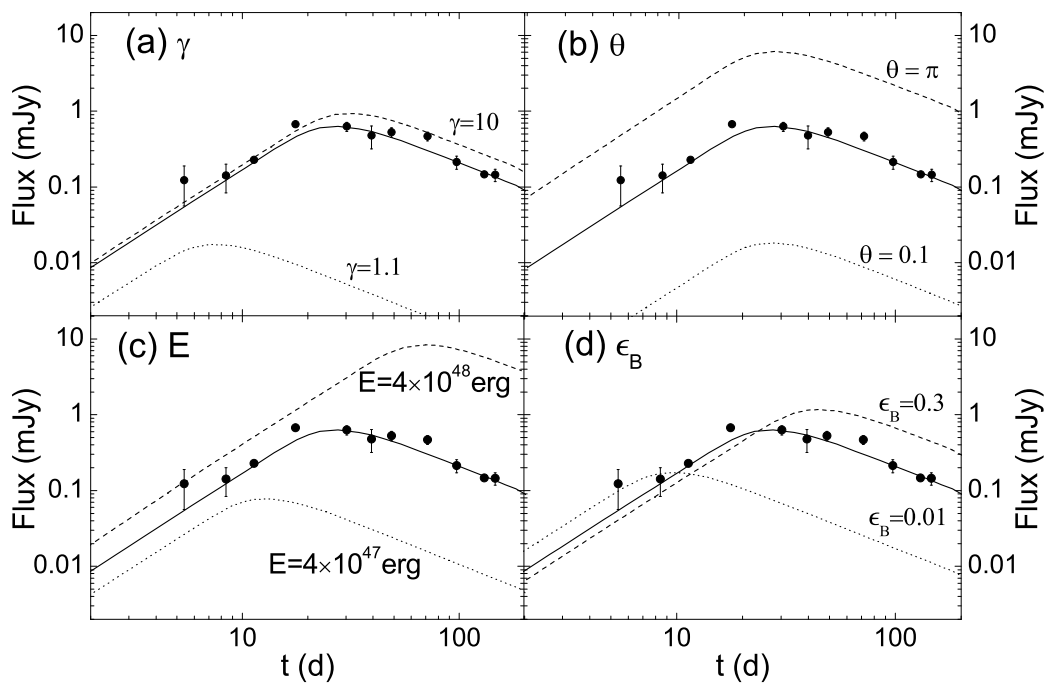}
\caption{Effects of the parameters $\gamma$, $\theta$, $E$, and $\epsilon_{\rm B}$ on the
1.4 GHz radio light curve. Observed data points of SN 2007gr are again taken from Soderberg et al. (2010b).
The solid curves are our best fit, while the dashed and dotted curves are light curves with only
one parameter altered, as marked in the panels.
\label{P1}}
\end{figure}

\begin{figure}
\epsscale{1.0}
\plotone{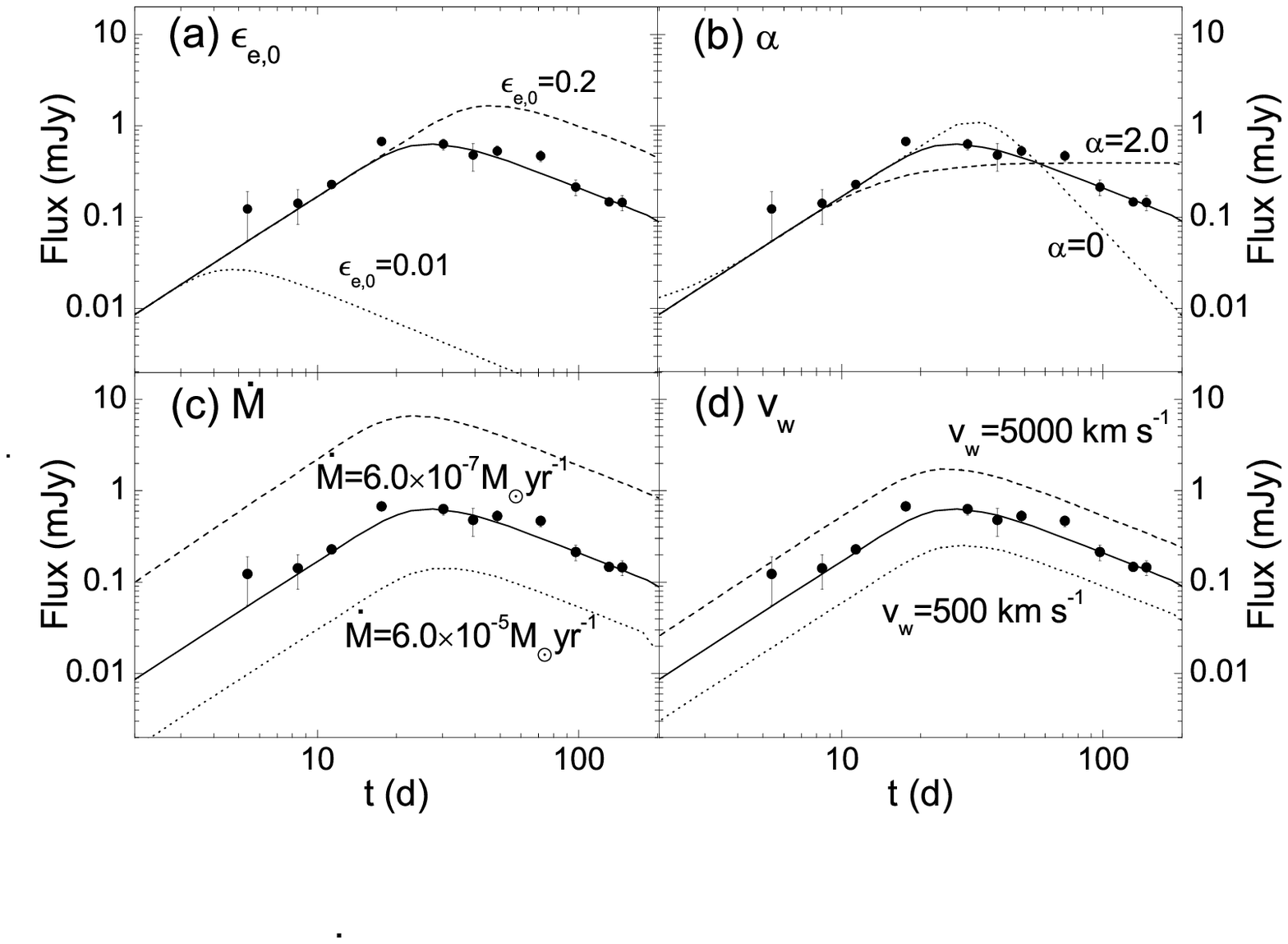}
\caption{Effects of the parameters $\epsilon_{\rm e}$, $\dot{M}$, and $v_{\rm w}$ on the
1.4 GHz radio light curve. Observed data points and the solid curves are the same as in
Fig. 3. The dashed and dotted curves are light curves with only
one parameter altered, as marked in the panels.
\label{fig4}}
\end{figure}

\section{Discussion and Conclusions}

In this paper, we propose an off-axis relativistic jet model to explain the multiband emission of
the Type Ic supernova SN 2007gr. From the observations of SN 2007gr by EVN (Paragi et al. 2010), we
adopted a model where the outflow contains a jet component and the
line of sight is off-axis.
The observing angle is
about $\theta_{\rm obs}=1.1$ rad.
We suggest that the radio emission come from
synchrotron radiation of relativistic electrons accelerated by
the shock produced by the collision between the jet and the
circumstellar medium.
Our calculations show that a jet with an initial half-opening angle of
$\theta_0 =0.6$ rad
and initial Lorentz factor of $\gamma_0=3$ can reproduce the emission
in radio band well. Optical and X-ray emission is also
consistent with the observational restrictions.

The jet engaged in our modeling of SN 2007gr differs significantly from
normal GRB jet. First, the Lorentz factor of normal GRB ejecta is typically
several hundred, while it is only a few here for SN 2007gr.
Second, the opening-angle of our SN jet
($0.6~\rm rad$) is much larger than that of a GRB jet ($\sim0.1~ \rm rad$).
Additionally, we would like to point out that the lateral expansion
speed of the jet is a relatively complicate factor. In our calculations, we
have assumed it to be the comoving sound speed. Although this is a reasonable
assumption, deviation may still be possible and the dynamics may be affected.
Third, the initial energy of our SN jet (isotropically $\sim10^{48}~\rm erg$) is much
smaller than that of GRB jet ($\sim10^{52}~\rm erg$).
We believe that the majority of energy is deposited into
supernova component in SN 2007gr explosion. Thus the above characteristics
are not too difficult to understand.

In view of Wolf-Rayet stars as the progenitors of type Ic SNe,
we assume the CSM as the typical wind of a Wolf-Rayet star with mass-loss rate
$\dot{M}=1.0\times10^{-5}~M_{\odot}~ \rm yr^{-1}$ and
$v_{\rm w}=1.5\times10^{3}~ \rm km~s^{-1}$.
A fast wind with low mass loss rate tend to increase the flux of the
radio emission.
The mass loss rate in our model is more than 10 times higher than that
in previous studies, but it is more typical for a Wolf-Rayet star.
Note that in our modeling, we have assumed that the CSM is purely
composed of protons and electrons. Actually, other nucleus such as
helium may also appear in the wind of Wolf-Rayet stars. Although their
presence in the CSM should not change the final results significantly,
we may still need to consider the factor in the future, when observations
become much improved.

For the electron energy faction ($\epsilon_{\rm e}$), we usually assume that it is constant
($\alpha=0$) in GRB model. However, in the current study,
a positive $\alpha$ is necessary in our fitting. It means that
$\epsilon_{\rm e}$ is increasing with time, i.e.,
the fraction of energy going to internal energy
increases with the deceleration of the shock.
Although the assumption of a varying $\epsilon_{\rm e}$ has also been
engaged in modeling some special GRBs and other transient objects (Rossi \&
Rees 2003; Ioka et al. 2006; Kong et al. 2010), note that the underlying physical
mechanism that leads to the variation of $\epsilon_{\rm e}$ is still much
uncertain.

The mutiband emission and spectra are quite different between SN 2007gr and normal GRBs.
As for GRBs, we can detect the prompt gamma-ray emission and multiband afterglows. The peak
frequency of prompt emission of GRBs is typically in gamma-ray band (about several hundred keV).
The peak frequency of GRB afterglows is typically in UV or X-ray band, and the X-ray afterglow
can usually be observed.
However, the emission from the jet of SN 2007gr is more like a failed GRB
rather than a normal GRB, since it is only a midly relativistic outflow.
It has been proposed that the prompt emission of a midly relativistic SN jet
come from the photosphere and bright in UV or soft X-ray band (Xu et al. 2010). The peak frequency
of afterglow should then be in near infrared or radio band. The X-ray emission
should be weak and hard to detect.

In the future, more and more relativistic SN ejecta will be detected.
These relativistic ejecta are considered as central engine driven.
In fact, supernova 2009bb was observed as another example of relativistic
supernova ejecta recently (Soderberg et al. 2010b).
If the very early emission of supernovae could be detected in the
future, it will be helpful for determining the speed of the
supernova ejecta directly.

\acknowledgments

We thank the anonymous referee for helpful comments and suggestions.
This work was supported
by the National Natural Science Foundation of China (Grant
No. 10625313 and 11033002),
the National Basic Research Program of China (973 Program, Grant
No. 2009CB824800) and the Grant-in-Aid for the `Global COE Bilateral
International Exchange Program' of Japan, Grant-in-Aid for Scientific
Research on Priority Areas No. 19047004 and Scientific Research on Innovative
Areas No. 21105509 by Ministry of Education, Culture, Sports,
Science and Technology (MEXT), Grant-in-Aid for Scientific Research (S)
No. 19104006 and Scientific Research (C) No. 21540404
by Japan Society for the Promotion of Science (JSPS).



\end{document}